\documentstyle[psfig]{l-aa}
\begin{document}

\author{L.H. Huang,  D.R. Jiang,  Xinwu Cao}
\title{A correlation analysis of flux ratios and the Doppler factor for
EGRET AGN sources
}
\date{ 
}

\thesaurus{(11.10.1; 11.11.1; 11.14.1)}
\offprints{D.R. Jiang(djiang@center.shao.ac.cn)}
\institute{ 
Shanghai Observatory,  Chinese Academy of Sciences,  80 Nandan Road, 
Shanghai,  200030,  P.R. China}

   \maketitle
   \markboth{L.H. Huang et al: A correlation analysis of flux ratios
   and the Doppler factor for EGRET AGN sources} {L.H. Huang et al: A correlation
   analysis of flux ratios and Doppler factor for EGRET AGN sources}

\begin{table*}[htb]
\caption[ ]{ Basic parameters of the selected AGNs (see text for details)
}
\begin{flushleft}
\renewcommand{\arraystretch}{1.2}
\begin{tabular}{llllllllllllll}\hline
\hline\noalign{\smallskip}

IAU NAME & class & redshift& $F_{\gamma}$ & $\alpha_{\gamma}$
& Ref. & $F_{opt}$ & Ref.& $F_{IR}$ & Ref.& $F_{90}$ & Ref.& $F_{230}$
&Ref. \\ 
~~~&~~~&~~~~&~~~~&~~~&~~~& mJy &~~~& mJy &~~~&
Jy&~~~&Jy~~~\\ \hline
\noalign{\smallskip}
0208-512 & HPQ & 1.003 & 110. & 0.7 & (14) & 0.66 & (6) &   &   &   &   &   &    \\
0234+285 & HPQ & 1.213 & 66.3 & 1.7 & (15) & 0.156 & (6) &   &   & 3.08 & (4) & 2.66 & (4)\\
0235+164 & B & 0.94 & 82.5 & 0.9 & (15) & 6.6 & (6) & 11.8 & (11) & 3.6 & (3) & 4.06 & (4)\\
0336-019 & HPQ &0.852&400 & 1.4&   (17)& 0.45& (7)& (1.1) & (11) &2.98 & (4)&1.38& (4)\\ 
0420-014 & HPQ & 0.915 & 61.2 & 0.9 & (15) & 0.296 & (6) & 7.0 & (10) & 5.52 & (3) & 4.49 & (4)\\  
0458-020 & HPQ & 2.286 & 31. & 1.5 & (17) & 0.17 & (6) &   &   & 2.16 & (4) & 0.71 & (3)\\
0521-365 & HPQ & 0.0554 & 20.7 & 1.2 & (15) & 2.0 & (6) & 19.5 & (12) & 4.63 & (3) & 3.24 & (4)\\  
0528+134 & LPQ & 2.06 & 295. & 1.6 & (15) & 0.062 & (6) & 0.65 & (18) & 3.48 & (1) & 2.1 & (1)\\
0537-441 & B & 0.894 & 36.4 & 1.0 & (15) & 2.05 & (6) & 10.0 & (12) &   &   &   &    \\
0716+714 & B & 0.3 & 50 & 1.4 & (14) & 20.5 & (6) & 11.1 & (9) & 2.75 & (3) & 3.03 & (4)\\  
0735+178 & B & 0.424 & 86.3 & 1.5 & (15) & 6.9 & (6) & 17.2 & (13) & 3.79 & (4) & 2.52 & (4)\\
0836+710 & LPQ & 2.172 & 45.3 & 1.4 & (15) & 0.98 & (6) &   &   & 0.86 & (2) & 0.56 & (2)\\
0954+658 & B & 0.368 & 14.3 & 0.9 & (15) & 0.82 & (7) &   &   & 0.58 & (2) & 0.65 & (2)\\
1101+384 & B & 0.031 & 21. & 0.9 & (14) & 17.8 & (6) & 50.1 & (11) & 0.68 & (1) & 0.5 & (1)\\
1127-145 & NP & 1.187 & 93.2 & 1.15 & (17) & 0.652 & (8) &   &   & 0.63 & (5) & 0.73 & (5)\\
1156+295 & HPQ & 0.729 & 229. & 1.0 & (17) & 5.1 & (6) & 1.91 & (13) & 2.44 & (4) & 1.18 & (3)\\
1219+285 & B & 0.102 & 32.2 & 0.3 & (16) & 2.9 & (7) & 10.6 & (13) & 1.63 & (2) & 1.51 & (2)\\
1226+023 & LPQ & 0.158 & 62.6 & 1.4 & (16) & 24.6 & (6) & 91.4 & (13) & 20.8 & (3) & 22.6 & (4)\\
1253-055 & HPQ & 0.538 & 450 & 0.9 & (17) & 15.1 & (6) & 21.9 & (11) & 19.1 & (4) & 19.6 & (4)\\
1510-089 & HPQ & 0.361 & 56.2 & 1.51 & (15) & 1.18 & (6) & 2.83 & (11) & 6.22 & (3) & 6.34 & (3)\\
1611+343 & LPQ & 1.40 & 54.9 & 1.0 & (15) & 0.39 & (6) & 0.68 & (1) & 1.8 & (1) & 1.19 & (1)\\
1633+382 & LPQ & 1.814 & 105.4 & 0.9 & (15) & 0.246 & (6) & 1.95 & (1) & 2.05 & (1) & 1.03 & (1)\\
1730-130 & LPQ & 0.902 & 136.9 & 1.4 & (15) & 0.52 & (8) &   &   & 9.85 & (4) & 4.98 & (3)\\
1739+522 & HPQ & 1.375 & 53.8 & 1.2 & (15) & 0.155 & (6) &   &   &   &   &   &    \\
2200+420 & B & 0.069 & 40 & 1.2 & (17) & 5.9 & (7) & 32.76 &(11) & 5.67 & (4) & 3.44 & (4)\\
2230+114 & HPQ & 1.037 & 28.5 & 1.6 & (15) & 0.47 & (6) & 1.45 & (11) & 3.73 & (4) & 1.97 & (4)\\
2251+158 & HPQ & 0.859 & 135 & 1.2 & (14) & 1.42 & (6) & 2.6 & (10) & 8.23 & (4) & 6.6 & (3)\\  
\noalign{\smallskip}
\hline
\end{tabular}
\renewcommand{\arraystretch}{1}
\end{flushleft}
\begin{list}{}{}
\item[(1)] Bloom et al., 1994
\item[(2)] Gear et al., 1994
\item[(3)] Steppe et al., 1988
\item[(4)] Steppe et al., 1992
\item[(5)] Steppe et al., 1995
\item[(6)] Dondi \& Ghisellini, 1995
\item[(7)] Ghisellini et al., 1993
\item[(8)] Comastri et al., 1996
\item[(9)] Sambruna et al., 1996
\item[(10)] Gear et al., 1985
\item[(11)] Mead et al., 1990
\item[(12)] Falomo et al., 1993
\item[(13)] Landau et al., 1986
\item[(14)] Fichtel et al., 1994
\item[(15)] Thompson et al., 1995
\item[(16)] Von Montigny et al., 1996
\item[(17)] Mattox et al., 1997
\item[(18)] Rieke et al., 1982
\end{list}
\end{table*}

\begin{table*}[htb]
\caption[ ]{VLBI and X-ray data of the selected AGNs 
}
\begin{flushleft}
\renewcommand{\arraystretch}{1.2}
\begin{tabular}{lllllllll}\hline\hline \noalign{\smallskip}
Source &  z & $\theta_d$ & $S_{c}(\nu_{s})$ & $\nu_s$
& Ref. & $S_{X}$ & Ref. & derived $\delta$\\
~~&~~~&mas& Jy & GHz& ~~& $\mu$ Jy & ~~&~~\\ \hline
\noalign{\smallskip}
0208-512 & 1.003 & 0.35 & 2.77 & 5.0 & (4) & 0.61 & (15) & 15.2\\
0234+285 & 1.213 & 0.09 & 1.7 & 22.3 & (5) & 0.15 & (5) & 16.6\\
0235+164 & 0.94 & 0.5 & 1.75 & 5.0 & (5) & 0.17 & (5) &6.5 \\
0336-019 & 0.852& 0.57&1.52 & 2.3 &  (5) & 0.047& (5)&15.6\\
0420-014 & 0.915 & 0.17 & 1.6 & 8.4 & (16) & 0.52 & (5)&14.0\\
0458-020 & 2.286 & 0.28 & 2.62 & 5.0 & (2) & 0.1 & (3) &47.0\\
0521-365 & 0.0554 & 0.73 & 1.82 & 5.0 & (4) & 0.68 & (5) &1.5\\
0528+134 & 2.06 & 0.08 & 3.0 & 22 & (8) & 1.59 & (10) &32.5\\  
0537-441 & 0.894 & 0.6 & 3.37 & 5 & (4) & 0.81 & (15) &6.8\\
0716+714 & $>0.3$ & 0.16 & 0.631 & 5.0 & (6) & 1.28 & (15)&7.0\\
0735+178 & 0.424 & 0.24 & 1.85 & 22.2 & (9) & 0.32 & (5)&2.0\\
0836+710 & 2.172 & 0.34 & 1.05 & 5.0 & (11) & 1.6 & (15)&8.0\\
0954+658 & 0.368 & 0.19 & 0.48 & 5.0 & (12) & 0.5 & (5)&5.1 \\
1101+384 & 0.031 & 0.24 & 0.366 & 5.0 & (6) & 14.0 & (5)&1.1 \\
1127-145 & 1.187 & 0.95 & 3.27 & 2.3 & (1) & 0.34 & (15)&12.0 \\
1156+295 & 0.729 & 0.123 & 1.4 & 22.2 & (5) & 0.15 & (5)&6.4\\
1219+285 & 0.102 & 0.20 & 0.159 & 5.0 & (13) & 0.42 & (5)&1.3 \\
1226+023 & 0.158 & 0.14 & 3.49 & 15. & (5) & 21 & (5)&6.0 \\
1253-055 & 0.538 & 0.14 & 4.84 & 15. & (5) & 1.4 & (5)&18. \\
1510-089 & 0.361 & 0.12 & 2.76 & 15 & (5) & 0.44 & (5)&14.4 \\
1611+343 & 1.40  &$< 0.38$ & 2.14 & 8.55 & (1) & 0.24 & (15)&7.1 \\
1633+382 & 1.814 & 0.5 & 5.4 & 10.7 & (14) & 0.08 & (3)&12.0 \\
1730-130 & 0.902 & 0.42 & 2.34 & 5 & (2) & 0.2 & (5)&11 \\
1739+522 & 1.375 & 0.37 & 0.89 & 5.0 & (5) & 0.1 & (5)&7.3\\
2200+420 & 0.069 & 0.35 & 1.6  & 5.0 & (5) & 0.82 & (5)&4.4 \\
2230+114 & 1.037 & 0.50 & 0.54 & 5.0 & (5) & 0.34 & (5)&1.9 \\
2251+158 & 0.859 & 0.54 & 5.226 & 5.0 & (7) & 5.5 & (3)&8.8 \\
\hline
\noalign{\smallskip}
\end{tabular}
\renewcommand{\arraystretch}{1}
\end{flushleft}
\begin{list}{}{}
\item[(1)] Fey et al., 1996
\item[(2)] Shen et al., 1997
\item[(3)] Dondi \& Ghisellini, 1995
\item[(4)] Shen et al., 1998
\item[(5)] Ghisellini et al., 1993
\item[(6)] Xu et al, 1995
\item[(7)] Cawthorne \& Gabuzda, 1996
\item[(8)] Pohl et al., 1995
\item[(9)] Baath et al., 1991
\item[(10)] Zhang et al., 1994
\item[(11)] Pearson \& Readhead, 1988
\item[(12)] Gabuzda et al., 1992
\item[(13)] Gabuzda et al., 1994
\item[(14)] Kellerman et al., 1977
\item[(15)] Comastri et al., 1997
\item[(16)] Wagner et al., 1995a
\end{list}
\end{table*}

\begin{abstract}
We present a correlation analysis between the flux ratio of high-energy
$\gamma$-ray emission
to synchrotron emission and the Doppler factor for a sample of EGRET AGNs.
The result favors a model that attributes the EGRET emission of AGNs to
inverse-Compton scattering on photons external to the jet.

\end{abstract}

\section{ Introduction}

  Since its launch in April 1991, the experiments on board of the
Compton Gamma Ray Observatory have detected many active
galactic nuclei (AGNs) at photon energies$>100$MeV. 42 AGNs have recently
been identified as EGRET sources with high confidence (Mattox 1997). A
common characteristic of these sources is that they all are radio-loud, flat
spectrum radio sources. Many of
them are seen as superluminal radio sources as well ($>25\%$). All these
AGNs belong to the blazar category (containing BL Lac objects, highly polarized (%
$>3\%$) quasars (HPQ), and optically violently variable (OVV) quasars), where
relativistic beaming is thought to play an important role in the $\gamma$%
-ray emission. In fact, strong evidence for relativistic motion of the
emitting plasma has been provided by the observation of superluminal
expansion of radio knots to the radio core in numerous blazars (Vermeulen \&
Cohen 1994). 

  Observationally, blazar energy spectra appear to be contained of at
least two components: a low-energy component with luminosity (${\nu}L_{\nu}$)
peaking in the IR-UV range, and a high-energy component with luminosity (${%
\nu}L_{\nu}$) strongly dominated by hard $\gamma$-rays, at least in those
sources detected by EGRET. The division of spectra into two components is
reflected by the deep drop of the low-energy component toward the far-UV
band (Impey \& Neugebauer 1988; Brown et al. 1989a, 1989b), and the rise of $%
{\nu}L_{\nu}$ toward higher energies in the X-ray band, observed clearly in
most OVV quasars (Worrall \& Wilkes 1990) and most radio selected BL Lac
objects (Sambruna et al. 1996). The simplest models addressing the
double-component nature of blazar
spectra are those in which the low-energy component is produced by
synchrotron radiation and the high-energy component is produced by the
inverse-Compton process. A variety of models have recently been proposed to
explain the origin of the $\gamma$-ray emission, including
synchrotron-self-Compton (SSC) radiation (Jones et al. 1974; Marscher 1980;
K$\ddot{o}$nigl 1981; Marscher \& Gear 1985; Ghisellini \& Maraschi 1989;
Maraschi et al. 1992) and inverse-Compton scattering on photons
produced by the accretion disc (Melia \& K$\ddot{o}$nigl 1989;
Dermer et al. 1992; Dermer \& Schlickeiser 1993), the
broad-line region (e.g. Sikora et al. 1994) or the dusty torus
(Wagner et al. 1995a). In addition to these models, there are also other
models devoted to understand the $\gamma$-ray emission mechanism of AGNs such
as: diffusive shock acceleration in electron-proton jets producing high-energy
emission through photomeson production and the subsequent cascade of
secondary particles (Mannheim \& Biermann 1992; Mannheim 1993a, 1993b);
synchrotron radiation in regions of high magnetic fields (Ghisellini 1993);
and radiation from secondary electrons resulting from decays of
ultrarelativistic neutrons emitted by the central
engine (e. g. Mastichiadis \& Protheroe 1990).

Considering the strong association of identified EGRET AGN
sources with the blazar class, the fact that nearly $25\%$ of the EGRET AGN sources
are superluminal, and the success of jet models in explaining the non-thermal
radiation from blazars, we will focus our discussion on $\gamma$-ray
emission models associated with jets. In fact, lately it has been
established for the blazars 0528+134 (Pohl et al. 1995), 3C279 (Wehrle et
al. 1996) and 1633+382 (Barthel et al. 1995) that enhanced levels of
activity in the optical and $\gamma$-ray bands are associated with the
emergence of new jet components. In several models that have been proposed,
the $\gamma$-ray emission originates in a jet as a product of inverse-Compton
scattering of relativistic electrons and seed photons produced externally to
the jet (Dermer et al. 1992; Sikora et al. 1994; Wagner et al. 1995a). The
importance of the process depends on the dominance of the externally
produced radiation energy density (as seen in the comoving frame of the jet)
with respect to magnetic energy density and the energy density of the
radiation produced internally. If the seed photons are produced externally to the
jet, then one can predict a correlation between the Doppler factor $\delta$
and the ratio of the $\gamma$-ray flux to the synchrotron radiation flux,
$({\nu}F_{\nu})_{\gamma}$/$({\nu}F_{\nu})_{syn}$, as was shown by Eq. (27)
in Dermer et al (1997):
\begin{equation}
\rho=\frac { ( {\nu} F_{\nu} )_{\gamma} } {({\nu}F_{\nu})_{syn}}
\approx k {\delta}^{1+\alpha}
\end{equation}
where $k=U_{iso}/U_{b}$ denotes the ratio of the energy density in the
external target radiation field and the blob's magnetic field.
The ratio of the SSC spectral power
flux at energy $\epsilon_{C}$ to the synchrotron spectral power flux at
energy $\epsilon_{s}$ is given by Eq. (28) in Dermer et al. (1997): 
\begin{equation}
\rho=\frac{2}{3}(\sigma_{T} n_{eo}
r_{b})(\frac {\epsilon_{s}}{\epsilon_{C}}) ^{\alpha-1}\ln
\Sigma_{C}(\epsilon_{C}) 
$$\rho_{SSC/syn}
\end{equation}
Where $r_{b}$ is the radius of the blob, assumed to be spherical
in the comoving frame, and $n_{eo}$ is the number density of nonthermal
electrons.
Therefore, the flux ratio in the SSC model is independent of the Doppler
factor. With the requirement that $\gamma$-rays do not interact with
X-rays (assumed cospatial) through photon-photon collisions, Dondi \&
Ghisellini calculated the lower limit on the Doppler boosting factor
$\delta$ of the $\gamma$
-ray emission for a sample of EGRET sources and found that
relativistic motion is required in all cases. They compared the derived
$\delta$ with other beaming factors such as the ratio between the $%
\gamma$-ray flux density and the optical flux density, and found no obvious
relation between the two different beaming factors. 

  The emission models mentioned above imply a variety of relations
between the emissions in different wavelength bands that can be used to
distinguish among them observationally. In this paper, we present
a compilation of flux densities in different wavelength bands for 27 blazars
which are
considered to be EGRET sources with high confidence (Mattox et al. 1997) and
the VLBI data of these sources if available. The data selection is described
in Sect. 2. In Sect. 3 we calculate the ratios of the $\gamma$-ray flux
density (here defined as ${\nu}F_{\nu}$) to the optical flux density,
the near-infrared (at 2.2 $\mu$m) flux density, and the millimeter (90 GHz, 230 GHz)
flux density. We derive the Doppler factor based on
the synchrotron-self-Compton model (Ghisellini et al. 1993), and we
investigate
the correlation between the flux ratio and the Doppler factor by using
a linear regression analysis. In Sect. 4, these correlation results are
compared with theoretical results and their physical meaning is discussed. 

\section{ Sample}

   Mattox et al. (1997) developed an analysis of EGRET radio
source identification that quantitatively incorporates a lot of important
information such as the size of the EGRET error region, the number density
of potentially confusing radio sources, the radio spectral index and the a
priori probability of detecting a radio source by EGRET. They provided a
table of 42 blazars that were considered to be robust identifications of
EGRET sources. In this paper, we will focus on a subset of these sources.
They are listed in Table 1, whose columns are: (1) IAU name; (2)
classification of the source (B=BL Lac object, HPQ=highly polarized quasar,
LPQ=low polarized quasar, NP=no polarization measurement); (3) redshift; (4) 
$\gamma $-ray flux density above 100 MeV, $F_\gamma $, in units of ${10}
^{-8}$ photon {cm}$^{-2}$ {s}$^{-1}$; (5) spectral index of the $\gamma $-ray
energy spectrum;
(6) reference for the $\gamma $-ray flux density and spectral index;
(7) V-band optical flux density $F_o$ in mJy; (8)
reference for the optical flux density; (9) near infrared flux density
$F_{IR}$ (at 2.2 $\mu$m)
in mJy; (10) reference for the infrared flux density; (11) flux
density at 90 GHz; (12) reference for the flux density at 90 GHz; (13) flux
density at 230 GHz; (14) reference for the flux density at 230 GHz. Table 1
lists 27 blazars which have been identified with high confidence by Mattox
et al (1997) and for which VLBI and X-ray observations (1 keV) are available.
All sources in Table 2 are known to be variable,
but only a few sources have been observed simultaneously
at different wavelengths
such as 3C273, 3C279. We therefore choose to list the highest flux densities
for all sources from the literature, irrespective of the dates of the observations.
The optical fluxes are mainly chosen from Dondi \& Ghisellini (1995) who
have chosen the maxima from the literature. The highest flux densities at
90 GHz and 230 GHz are mainly chosen from Steppe et al. (1988, 1992). The
classifications mainly refer to Dondi \& Ghisellini (1995) and
Ghisellini et al. (1993). Table 2 lists the VLBI and X-ray observations;
(1) IAU name; (2) redshift; (3) VLBI core size($\theta _d
$) in mas; (4) core radio flux density ($F_c$) at frequency $\nu _s$; (5)
observation frequency $\nu _s$ in GHz; (6) reference for the VLBI data;
(7) 1 keV X-ray flux density $F_X$ in $\mu $Jy; (8) reference for the X-ray flux.
The X-ray flux densities are mainly chosen from Ghisellini (1993). For those
sources which have multi-frequency VLBI observations, the VLBI data at the
highest frequency are chosen.

\section{ Results}

\subsection{   The derivation of the Doppler factor}

  By comparing the predicted and observed self-Compton flux, one can
derive $\delta$ in the case of a moving sphere (p=3+$\alpha$) (Ghisellini et
al. 1993)

\begin{equation}
\delta=f(\alpha) F_{c} \left[ {\frac{{\ln ({\nu}_{b}/{\nu}_{m})} }{{\ F_{x} {
\theta_{d}}^{6+4\alpha} {{\nu}_{x}}^{\alpha} {{\nu}_{c}}^{5+3\alpha}} }}
\right]^{1 / {(4+2\alpha)}} 
\end{equation}
where $F_{m}$ and $F_{x}$ are the radio flux density and X-ray flux density
in Jy, ${\nu}_{x}$ is in keV, ${\nu}_{c}$ is the frequency of VLBI observation
in GHz, $
\theta_{d}$ is in mas, ${\nu} _{b}$ is the synchrotron high frequency
cutoff (assumed to be $10^{14}$ Hz), and the function
$f(\alpha)=0.08{\alpha}+0.14$ (here $\alpha=0.75$ is assumed).
Recently, an improved approach was
proposed by Jiang et al. (1998) to derive the Doppler factors based on an
inhomogeneous jet model. However, the proper motion of knots is required in
their derivation, which limits the number of sources in the sample. To make
the sample as large as possible for the statistical analysis, we use the
spherical model in this paper. The derived Doppler factors $\delta$ are
listed in Table 2. 

\subsection{{   Correlation between $\rho_{opt}$ and $\delta$, $ 
\rho_{NIR}$ and $\delta$}}

  Also for the $\gamma $-ray flux densities of the EGRET sources, the
maximum observed fluxes ($>100$MeV) are chosen.
The $\gamma $ -ray photon flux of  is transferred into an integrated energy
flux between 0.1 GeV
and 5 GeV (i.e. ${\nu }_\gamma F_\gamma $). The reasons for choosing the
maximum fluxes are: (1) all AGNs are variable sources and their radio
emissions are mainly produced via synchrotron emission; (2) in the optical band,
both the blue bump and the starlight account for part of the optical
emission, the synchrotron radiation varies rapidly and the contribution of
both the blue bump and the starlight can be reduced by using the maximum
flux; (3) also, for the emission in the near infrared band, the contribution of
thermal emission coming from the torus can be reduced. Figs. 1 and 2 show the
ratio of the $\gamma $-ray flux to the optical flux versus the Doppler factor, and
the ratio of the $\gamma $-ray flux to the infrared flux versus the
Doppler factor, respectively. There is an obvious correlation between the
flux density ratio and the Doppler factor.
The relation between
the flux ratio and the Doppler factor is: 
\begin{equation}
\frac{(\nu F_\nu )_\gamma }{(\nu F_\nu )_{opt}}\propto \delta ^{1.33}, 
\end{equation}
and 
\begin{equation}
\frac{(\nu F_\nu )_\gamma }{(\nu F_\nu )_{NIR}}\propto \delta ^{1.34}. 
\end{equation}
When BL Lac objects are
excluded,
the correlation remains present, but is less significant. 

\subsection{{   Correlation between $\rho_{mm}$ and $\delta$}}

  Figs. 3 and 4 show the ratio of the $\gamma$-ray flux density to
the millimeter
(90 GHz, 230 GHz) flux density versus the Doppler factor. The correlation
between
$\rho_{mm}$ and $\delta$ is poor. They
improve when BL Lac objects are excluded, but are still
poor in comparison to the optical and near-infrared cases.

   \begin{figure}
 \centerline{\psfig{figure=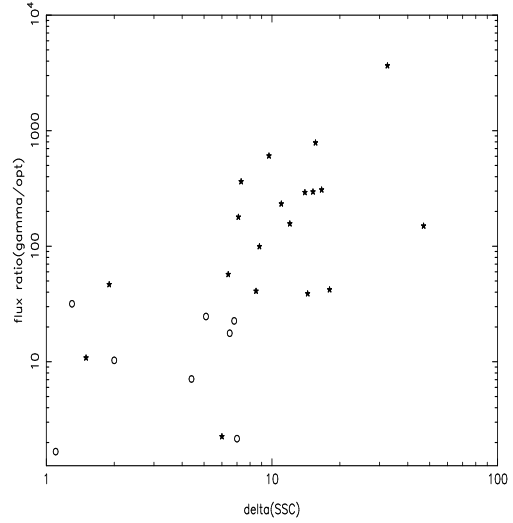,width=8.0cm,height=8.0cm,angle=-90}}
     \caption{
Ratio of $\gamma$-ray flux to optical flux vs. Doppler factor $\delta$.
Circles: BL Lacs; Stars: quasars.}
         \label{}
   \end{figure}

   \begin{figure}
 \centerline{\psfig{figure=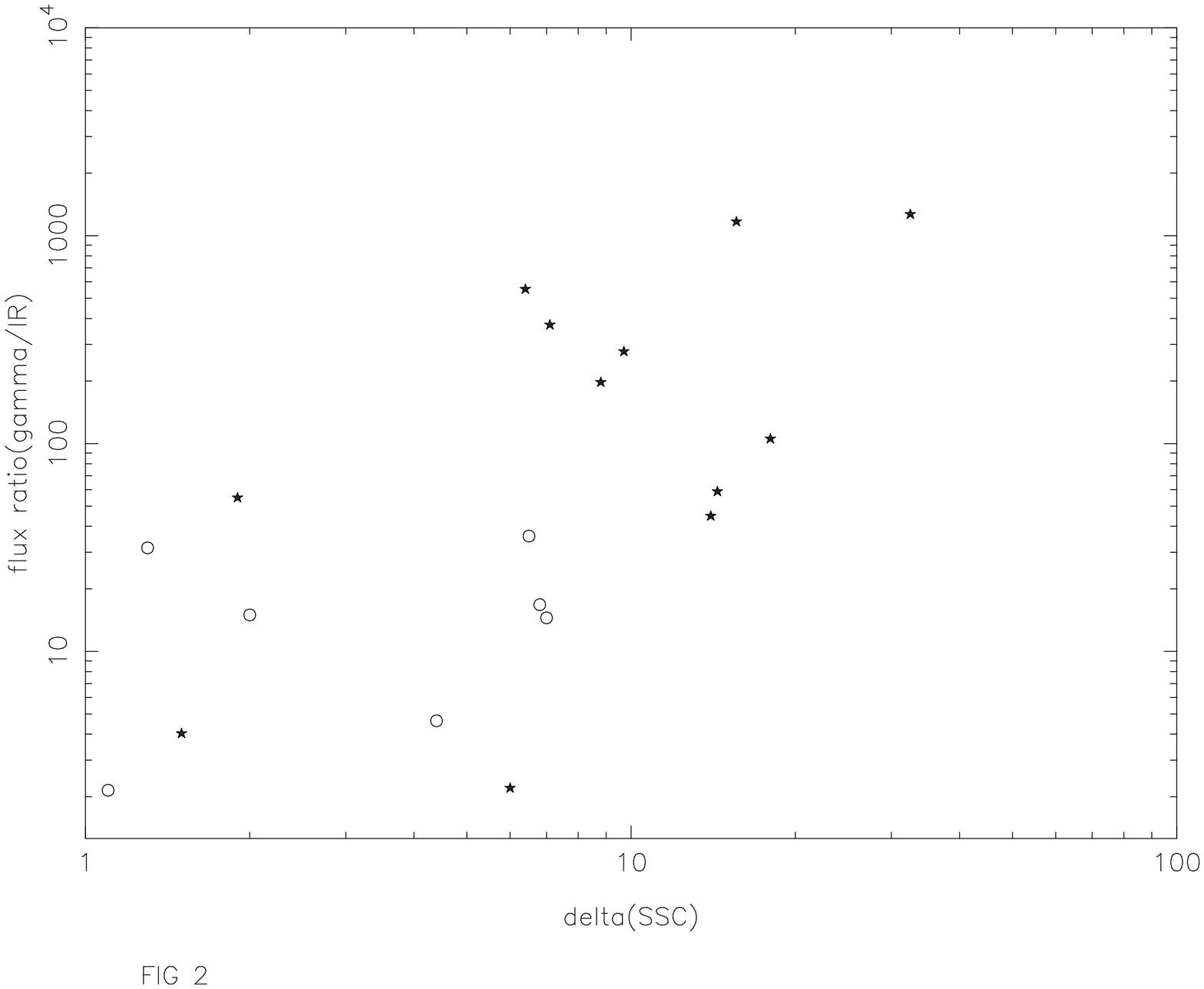,width=8.0cm,height=8.0cm,angle=-90}}
     \caption{
Same as Fig. 1, but for the ratio of $\gamma$-ray flux to near infrared
flux
}
         \label{}
   \end{figure}

   \begin{figure}
 \centerline{\psfig{figure=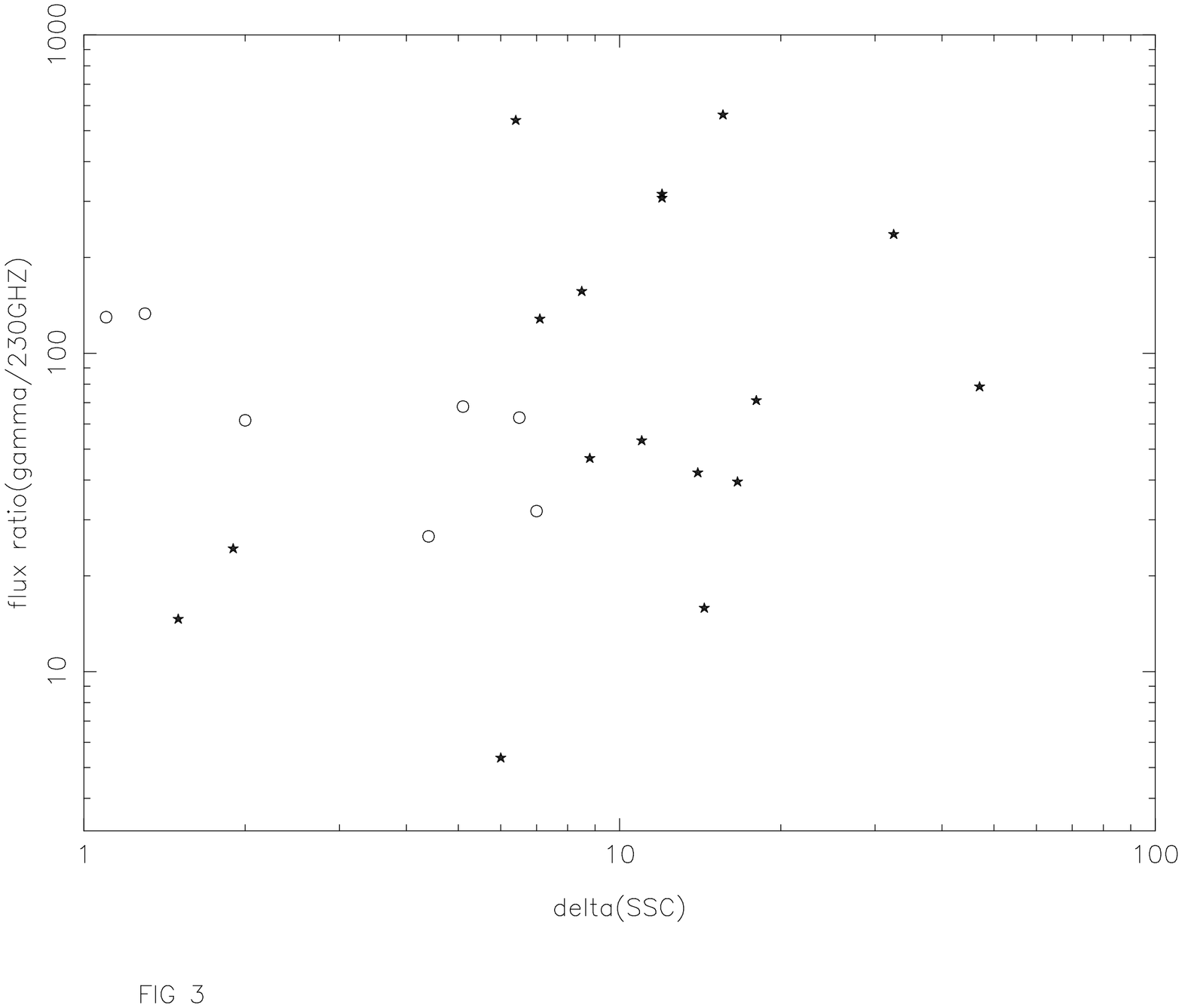,width=8.0cm,height=8.0cm,angle=-90}}
     \caption{
Same as Fig. 1, but for the ratio of $\gamma$-ray flux to the flux at
230 GHz
   }
         \label{}
   \end{figure}

   \begin{figure}
 \centerline{\psfig{figure=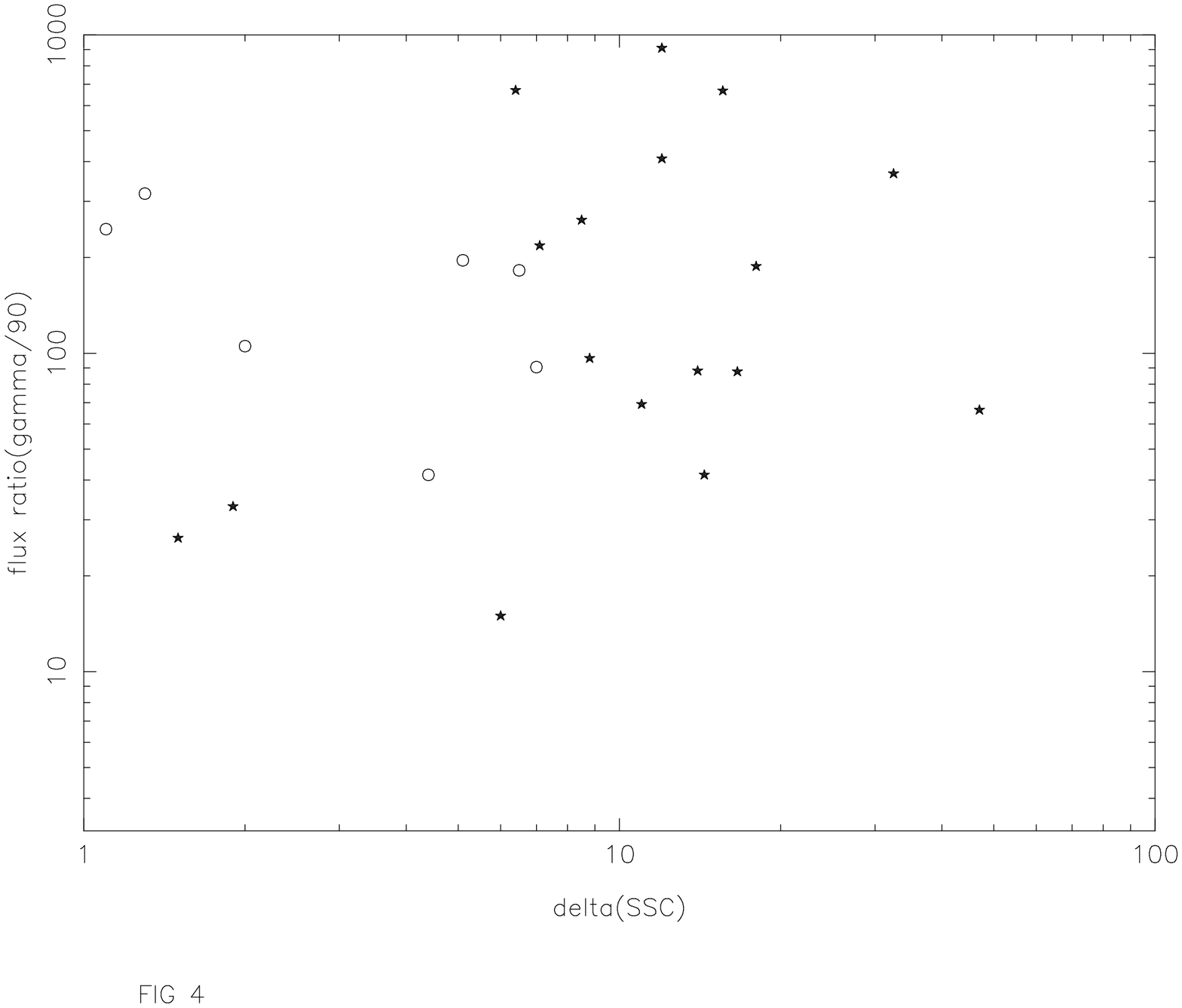,width=8.0cm,height=8.0cm,angle=-90}}
     \caption{
Same as Fig. 1, but for the ratio of $\gamma$-ray flux to the flux at
90 GHz}
         \label{}
   \end{figure}

\begin{table*}[htb]
\caption[ ]{Results of the linear regression analysis 
}
\begin{flushleft}
\renewcommand{\arraystretch}{1.2}
\begin{tabular}{llllllll}\hline\hline\noalign{\smallskip}
y & x & N & a & b & r & P & note.\\ \hline
\noalign{\smallskip}
log$\frac{(\nu F_{\nu})_{\gamma}}{(\nu F_{\nu})_{opt}}$ & log($\delta$)
&27 & 0.61 & 1.33 & 0.66 & $2\times{10}^{-4}$ &   \\
   &   
&19& 0.94 & 1.15 & 0.57 & $1.05\times{10}^{-2}$ & excluding BL Lac objects\\
log$\frac{(\nu F_{\nu})_{\gamma}}{(\nu F_{\nu})_{NIR}}$ & log($\delta$)
&19 & 0.64 & 1.34 & 0.65 & $2.5\times{10}^{-3}$ &   \\
   &
&12 & 0.8 & 1.32 & 0.59 & $4.5\times{10}^{-2}$ & excluding BL Lac objects  \\

log$\frac{(\nu F_{\nu})_{\gamma}}{(\nu F_{\nu})_{230GHZ}}$ & log($\delta$)
&24 & 1.6 & 0.26 & 0.22 & 0.3 &   \\
   &   
&17 & 1.3 & 0.58 & 0.14 & 0.14 & excluding BL Lac objects\\
log$\frac{(\nu F_{\nu})_{\gamma}}{(\nu F_{\nu})_{90GHZ}}$ & log($\delta$)
&24 & 2.0 & 0.19 & 0.12 & 0.4 &   \\
    &
&17 & 1.59 & 0.53 & 0.37 &  0.14 & excluding BL LAC objects\\
\hline
\noalign{\smallskip}
\end{tabular}
\renewcommand{\arraystretch}{1}
\end{flushleft}
\begin{list}{}{}

\item[Notes:] The linear regression is obtained by considering x to be the
independent variable and assuming a relation y=a+bx; N is the number of points,
r is the correlation coefficient, and P is the chance probability.

\end{list}

\end{table*}

\section{   Discussion}

  That the soft photons are Compton up scattered to the $\gamma $-ray
range by nonthermal electrons in the relativisticly moving blobs might be the
most favorable model for the $\gamma $-ray emission. The soft photons could
come from synchrotron emission in the blob, namely the SSC process, or from
outside the moving blob.
Dermer (1995) proposed
that the observed optically thin synchrotron flux density, combined with the
Compton-scattered photon flux density, can be used to test whether the
high-energy radiation is produced through SSC or external Compton 
scattering. We have examined the
correlations between the ratio of the $\gamma $-ray flux density to the flux
density in different wavebands and the Doppler factor $\delta $. Two
significant correlations are found, namely between $\rho _{opt}$and $\delta $,
and between $\rho
_{NIR}$ and $\delta $, which support the external Compton
scattering models for the high-energy $\gamma $-ray emission. The obtained
correlations suggest that the observed flux densities in the optical and
near-infrared wavebands are mainly due to optically thin
synchrotron emission in the blobs. The relation between the emission in the
optical band and the emission in the $\gamma $-ray band was also shown by the
facts that the highest $\gamma $-ray flux densities were recorded during the
most prominent flare in the optical range for PKS 0420-014 (Wagner et al.
1995a) and that a rapid flare peaked about 22 hours after the optical outburst
for PKS 1406-076 (Wagner et al. 1995b).

  Table 3 also lists the values of b, which corresponds to $1+\alpha$
in formula (1). The values of b are indeed close to $1+\alpha$ ($\alpha=0. 75$ has
been assumed), in the optical case and in the infrared case ($b=1.33$ and $ 
b=1.34$, respectively). The k in formula (1) is the ratio of the isotropic
energy density to the energy density of magnetic field, which is about 4-5 in the optical and
NIR cases. 

  The Doppler factors $\delta$ of the BL Lac objects derived here are
lower than those of other sources in the sample. The mechanism for X-ray
radiation of BL Lac objects is still not quite clear. We used here a
single model to derive the Doppler factor $\delta$ for all sources in the
sample, including BL Lacs, on the assumption that SSC accounts for X-ray
emission. However, some authors (Urry 1994) suggested that the X-ray
emission from BL Lacs is due to synchrotron radiation instead of the
synchrotron self-Compton radiation. If this is true, the derivation of the
BL Lacs' Doppler factor $\delta$ here is questionable. 

  Compared with the optical and NIR cases, the correlation $ 
\rho_{mm}-\delta$ is quite poor. One
possible reason is that the synchrotron radiations in the millimeter wavebands
is optically thick for some sources in the sample, whereas the theoretical
model requires an optically thin synchrotron flux density. In additions, the
correlations between the flux ratios and the Doppler factor show
some deviation from the theoretical value b= $1+\alpha$. This may not be 
surprising since
there are several factors that can affect the flux ratio $\rho$. For
example, the SSC radiation may account for part of the
$\gamma$-ray emission as well. For a few sources such as Mrk421, strong
radiation is seen at 
TeV energies, and for others such as 3C273, the $\gamma$-ray emission
is probably concentrated in MeV band, which will introduce errors in the
correlation as well. In addition, the Doppler factor $\delta$ we have derived
is based on the blob model. To obtain this result, one
must know the core angular size and the flux density at the turnover
frequency. In practice, it is difficult to obtain this information, so one
has to assume the VLBI observing frequency to be the synchrotron
self-absorption frequency. Furthermore, a single $\alpha=0.75$ is assumed for
all sources in our sample. The derivation of the
Doppler factor based on the blob model is clearly simplified. 

  By assuming that X-rays and $\gamma$-rays are cospatial, and
deriving the size of the source using the variability time-scale, Dondi \&
Ghisellini (1993) derived a lower limit $\delta_{\gamma}$ on the Doppler
factor of a source. They compared the Doppler factor with the flux
ratio between the $\gamma$-ray flux density and the optical flux density and
found a marginal correlation. They also listed the Doppler factors derived
from the SSC model, but they did not compare the flux ratio with the $ 
\delta_{SSC}$. We compare their flux ratio with their $\delta_{SSC}$ and
also find a strong correlation when excluding sources which were
not identified with high confidence by Mattox et al (1997).

  The strong correlations between
$({\nu}F_{\nu})_{\gamma}$/$({\nu}F_{\nu})_{opt}$
and $\delta$ as well as
$({\nu}F_{\nu})_{\gamma}$/$({\nu}F_{\nu})_{IR}$
and $\delta$ show that the $\gamma$-ray emission is more related to emission
in the NIR-OPT band than to emission in the millimeter band. These results
suggest that
the relativistic electrons which are responsible for the optical emission are
responsible for the $\gamma$-ray emission as well. This favors a model in which
the $\gamma$-ray emission is due to inverse-Compton scattering on
photons external to the jet.

\section{ Conclusion}

  We have studied a sample of EGRET AGNs which have recently been identified
with high confidence and derive the Doppler factors for most of these gamma
sources by using VLBI and X-ray data on the assumption of a homogeneous
spherical emission plasma. Strong correlations between the flux ratio $
\frac{(\nu F_{\nu})_{\gamma}}{(\nu F_{\nu})_{opt}}$ and $\delta $ as well as
$\frac{(\nu F_{\nu})_{\gamma}}{(\nu F_{\nu})_{IR}}$ and $\delta $ are found,
which is expected for inverse-Compton scattering on photons from outside the
jet. The poor correlations between the flux ratio $\frac{(\nu
F_{\nu})_{\gamma}}{(\nu F_{\nu})_{mm}}$ and $\delta $ may be due to the
optically thick synchrotron emission in the millimeter wavelength for some
EGRET sources.

\begin{acknowledgements}
We thank the referee, H. Bloemen, for his helpful comments and the linguistic improvements
on the manuscript.
The support from Pandeng Plan and NNSFC (No. 19703002) is gratefully
acknowledged.
\end{acknowledgements}

\end{document}